\documentclass{iopart}             
\usepackage{iopams}                
\usepackage{epsfig}
\def\rma{{\rm a}}
\def\rmd{\,{\rm d}}
\def\rme{{\rm e}}
\def\rmp{{\rm p}}
\def\rmr{{\rm r}}
\def\rms{{\rm s}}
\def\rmcm{{\rm cm}}

\def\rmC{{\rm c}}

\def\kb{k_{\rm B}}

\def\ell{l}   
\def\be{\begin{equation}}
\def\ee{\end{equation}}
\def\AA{{\it Astron. Astrophys. }}

\def\APJ{{\it Astrophys. J. }}
\def\APJS{{\it Astrophys. J. Suppl. Ser. }}
\def\CPC{{\it Comput. Phys. Commun. }}
\def\JQSRT{{\it J. Quant. Spect. Rad. Trans. }}
\def\MNRAS{{\it Mon. Not. R. astr. Soc. }}

\begin{document}

\setlength{\arraycolsep}{2.5pt}             
\jl{2}
\setcounter{footnote}{2}
\title[Inner-shell contributions to opacity]
{On the importance of inner-shell transitions for opacity calculations}
\author{N R Badnell$^1$ and M J Seaton$^2$}
\address{$^1$Department of Physics, University of
             Strathclyde, Glasgow G4 0NG, UK}
\address{$^2$Department of Physics and Astronomy, University College
   London, London WC1E 6BT, UK}
\begin{abstract}
For high temperatures and densities, stellar opacities obtained from the
Opacity Project (OP) were smaller than those obtained from the OPAL project.
Iglesias and Rogers [\APJ  {\bf 443} 469 (1995)] suggested that the 
discrepancy was due to the omission by OP of important atomic inner-shell
processes, and considered in detail results for a mixture of 6 elements:
H, He, C, O, S and Fe. Extensive new inner-shell data have now been
computed using the code {\sc autostructure}. It is shown that the inclusion
of these data in the OP work gives opacities for the 6-element mix 
which are in much closer agreement with those from OPAL. We also 
discuss a number of problems relating to the calculation of
opacities and of equations-of-state in dense plasmas.
\end{abstract}
\submitted
\section{Introduction}
Energy is produced by nuclear reactions at the centre of a star,
at a temperature of a few  times $10^7$ K, and escapes at the
stellar surface. The structure of a star is
determined by the equations for conservation of mass and of energy,
an equation for  
hydrostatic equilibrium, and by the temperature gradient (see
\cite{S,CG,C}). In regions for which
convection does not occur, the temperature gradient is determined
by the {\it Rosseland-mean opacity}, which is the concern of 
the present paper. 

In a stellar interior, an atom of a chemical element $k$ can exist in
a number of ionization stages, $i$, and energy levels, $j$. Using 
the frequency variable $u=h\nu/(\kb T)$ (where $\kb$ is the Boltzmann
constant), let
$\sigma_{ijk}(u)$ be the cross section for absorption\footnote{For absorption
processes, the correction factor for stimulated emission,
$[1-\exp(-u)]$, is included.} or scattering
of radiation by level $(ijk)$ and let $p_{ijk}$ be the probability of
that level being populated. Then, the opacity cross-section for 
element $k$ is
\be 
\sigma_k(u)=\sum_{i,j}p_{ijk}\sigma_{ijk}(u)\,.
\ee
With $f_{k}$ the fractional abundance for element $k$, normalized to
\be 
\sum_k f_k=1\, ,
\ee
the mean cross-section per atom for a mixture of chemical elements is
\be 
\sigma(u)=\sum_k f_k\sigma_k(u)\,.
\ee
The {\it Rosseland-mean cross-section} is $\sigma_{\rm R}$, where
\be 
\frac{1}{\sigma_{\rm R}}=\int_0^\infty \frac{1}{\sigma(u)} F(u) \rmd u 
\ee
and 
\be 
F(u)=[15/(4\pi^4)]u^4\exp(-u)/[1-\exp(-u)]^2 
\ee
(see \cite{S,CG,C}).
Astronomers usually use opacities {\it per unit mass}. The Rosseland-mean
opacity per unit mass is
\be 
\kappa_{\rm R}=\sigma_{\rm R}/\mu \, ,
\ee
where $\mu$ is the mean atomic weight.

The  1982 paper of  Simon \cite{sim} suggested that possible
errors in opacities  might
explain discrepancies between theory and observations for pulsational
properties of stars, and
provided the stimulus for two major new efforts in opacity calculations:
one referred to as OPAL at the Lawrence Livermore National Laboratory by
C.~A.~Iglesias and and F.~J.~Rogers; and the other referred to as The Opacity
Project, OP. Early OPAL results \cite{IRW} showed that the inclusion of
very large numbers of spectrum lines did indeed lead to major revisions
in opacities. A number of subsequent OPAL papers have been published,
of which one of the most recent is \cite{IR96}. 
The present paper is a continuation of the OP work. The calculations of 
atomic data were described in a series of
papers in this journal, of which the first was \cite{ADOC1}, and
OP results for opacities were published in \cite{symp}, to be referred to as 
SYMP. A collection of papers from the OP work, together with selected
atomic data tables, have been published in book form \cite{team}.

For regions of stellar interiors with temperatures of a few
times $10^5$ K, which are of particular importance for pulsation
studies, opacities from OPAL and OP are in close agreement and
can be larger than values previously adopted, by up to factors
of about 3. The new opacities have led to substantial improvements
in the agreement between calculated and observed pulsation properties.
However, for deeper layers, the results from OPAL were larger
than those from OP, by amounts of 30\% or so.

Very large numbers of spectrum lines are included in the calculations,
of order a few times $10^7$ for a complete opacity run. OPAL used a parametric
potential model \cite{RWI}, while OP (see \cite{team}) used $R$-matrix calculations
supplemented by data from Kurucz \cite{K} and from {\sc superstructure} 
\cite{PLUS}. Much of the OP data includes allowance for configuration-interaction
effects which are not included in the OPAL work.
In their 1995 paper \cite{IR95}, to be referred to as IR95,
Iglesias  and Rogers suggested that the discrepancies
between OPAL and OP at the higher temperatures and densities occurring
in the deeper layers of stellar interiors could be due to the 
omission by OP
of some important inner-shell transitions.
We have now made new calculations, using the code {\sc autostructure} \cite{AS1},
for promotions of inner-shell electrons, both via photoionization and
photoexcitation of autoionizing states. Our results confirm the essential
correctness of the suggestion made in IR95. The present paper
describes the new atomic-physics calculations and presents results for 
the 6-element mix of IR95.

The structure of the paper is as follows: in section 2 we discuss issues relating
to the equation-of-state; in section 3 we detail the new inner-shell transitions
that we now include
and the methodology that we used to describe them; in section 4 we briefly discuss free--free
transitions: in section 5 we present and discuss our results for Rosseland-mean opacities;
in section 6 we discuss a number of issues that arise in determining opacities;
in section 7 we look at the opacity in the solar centre region; and finally, 
in section 8, we give a brief summary.
\section{The equation-of-state (EOS)}
The fraction of element $k$ in ionization stage $i$ is
\be 
\phi_{ik}=\sum_j p_{ijk}\,.
\ee
We may put
\be 
p_{ijk}=\phi_{ik}\times g_{ijk}W_{ijk}\exp[-E_{ijk}/(\kb T)]/U_{ik} 
\ee
where, for level $(ijk)$, $g_{ijk}$ is statistical weight, $E_{ijk}$
the total energy and $W_{ijk}$ an {\em occupation probability}.
The {\em internal partition function} is
\be 
U_{ik}=\sum_j g_{ijk}W_{ijk}\exp[-E_{ijk}/(\kb T)]\,.
\ee
If all $W_{ijk}$ were set to unity, the summation for $U_{ik}$ would
be divergent. Due to interactions with particles and fields of the
surrounding plasma, states of sufficiently high energy have only
small probabilities of being occupied and, hence, small values of $W_{ijk}$.
\subsection{Occupation probabilities}
The methods used in the OP work for the calculation of the $W_{ijk}$ are 
described by Hummer and Mihalas in \cite{hm}. Let $F$ be the ion
micro-field and $P(F)$ the micro-field distribution (MFD): 
$P(F)\rmd F$ is the probability of $F$ being in
the range $\rmd F$, with $\int_0^\infty P(F)\rmd F =1$.
Hummer and Mihalas
define critical fields, $F_{ijk}$, such that species $(ijk)$ can
only exist in an environment with $F<F_{ijk}$, giving
\be 
W_{ijk}=\int_0^{F_{ijk}} P(F) \rmd F\,. 
\label{10}
\ee
The use of equation (\ref{10}) gives $W\simeq 1$ for low densities 
but $W$ to be small for high states and high densities.
In \cite{hm} Hummer and Mihalas used the Holtsmark MFD,
which does not make full allowance for correlations
between particles nor for the charge on the ion $(ijk)$. They also
introduced a simple analytical approximation (equation (4.70) of
\cite{hm}) which gave results in close agreement with those from
(\ref{10}) at low densities  but a more rapid exponential
decrease at high densities. A similar 
exponential form is obtained using a `hard-sphere' approximation
(see section III (a) of \cite{hm}). The analytical form, (4.70) of
\cite{hm}, was  used in SYMP.

Iglesias and Rogers, in IR95, noted that the Holtsmark distribution
would not be a good approximation at the higher densities of stellar
interiors and Nayfonov {\em et al.} \cite{nay} subsequently obtained improved
expressions for the $W_{ijk}$ using the micro-field distributions of
Hooper \cite{hoop}, which are also used in the OP calculations of
line-profiles for hydrogenic ions \cite{adoc13}. Following 
Nayfonov {\em et al.}, we refer to the equation of state using their
MFD as the Q-EOS. Their $W_{ijk}$ 
are fairly close to those obtained using the APEX
distribution of IR95 and, with one modification to be described  
below, will be used in the present work.
\subsection{The ionization equilibrium} 
For simplicity of presentation,  
we here omit specification of the element index $k$ and
neglect all refinements such as 
allowance for electron degeneracy. The conclusions reached remain
valid when all necessary refinements are included.
We take the ionization index $i$ to be equal to the number of
bound electrons: $i=0$ for the bare nucleus  and $i=Z$ for a neutral
atom with nuclear charge $Z$.

The ratio of ionization fractions in successive stages is (see \cite{ADOC1}
or \cite{hm})
\be 
\frac{\phi_i}{\phi_{i-1}}=\frac{U_i}{U_{i-1}}\times \frac{N_\rme}{U_\rme}\,, 
\ee
where $N_\rme$ is the electron density, 
\be 
U_\rme=2\left\{\frac{m_\rme\kb T}{2\pi\hbar^2}\right\}^{3/2} 
\ee
and $m_\rme$ is the mass of an electron.

Since $U_0=1$, the fraction in stage
$i$ relative to the fraction of bare nuclei is
\be 
\frac{\phi_i}{\phi_0}=U_i\times \left[\frac{N_\rme}{U_\rme}\right]^i. 
\label{14}
\ee
\subsection{Pressure ionization}
Now consider fixed $T$ and increasing $N_\rme$. Initially,
$\phi_i/\phi_0$ will increase with increasing $N_\rme$ due to the
factor  $[N_\rme/U_\rme]^i$ in (\ref{14}). That effect is {\em pressure 
recombination}. Eventually, for large densities,
the $W_{ij}$ in $U_i$ become small and, if they become sufficiently small,
pressure recombination can be followed by {\em pressure ionization}.

For  $W$ calculated using (\ref{10}), in the 
limit of $N_\rme$ large one obtains $W\propto N_\rme^{-2}$ using the
Holtsmark MFD and $W\propto N_\rme^{-3/2}$ with the Q form. In 
either case it is seen that pressure ionization does not 
occur for $i\geq 2$. That result would appear to be quite 
unphysical, since it would imply that all atoms are in 
states which have very small occupation probabilities. Pressure
ionization does occur, for all $i$,
if one uses a `hard-sphere' model
for $W$ or, as in SYMP,  the approximation of equation (4.70) of \cite{hm}, 
giving exponential decreases of $W$ with increasing $N_\rme$.

We conclude that equation (\ref{10}) can be expected to give reasonable
values for $W$ when $W$ is not very small, but that it does not give
a sufficiently rapid decrease of $W$ in the limit of high
densities.
We adopt the expedient of introducing a critical value $W_\rmC$ of
$W$; use the value of $W$ from equation (\ref{10}) if it is greater than
$W_\rmC$; and take $W=0$ if equation (\ref{10}) gives $W<W_\rmC$. For quite
a wide range of values of $W_\rmC$, results for opacities are
found to be insensitive to the value of $W_\rmC$ adopted. The final results
reported in the present paper are obtained with $W_\rmC=10^{-3}$.

The need for introducing a cut-off in $W$ is illustrated in Figure~1
for carbon at rather high temperature ($\log(T)=7.5$) and very high
densities ($\log(N_\rme)=27.0$, 27.5 and 28.0). Figure~1~(a) shows
ground-state occupation probabilities $W$ against ionization stage
$i$. Neutral carbon, $i=6$, is seen to have very small values of 
$W$ which decrease with increasing $N_\rme$. Figure 1 (b) shows ionization
fractions calculated without a cut-off in $W$. There is seen to be
an abrupt change from the case of $\log(N_\rme)=27.0$  with carbon nearly
fully-ionized (ionization fraction 0.98 for $i=0$) to the case
of $\log(N_\rme)=28.0$ with carbon nearly fully-neutral (ionization
fraction 0.95 for $i=6$). The result for $\log(N_\rme)=28$ is clearly
nonsensical, with nearly all of the carbon in a state with occupation
probability of $W\simeq 10^{-12}$ ! With a cut-off of $W_\rmC=10^{-3}$ we 
obtain a result which is much more plausible: for $\log(N_\rme)=27.0$,
ionization fractions very similar to those for $W_\rmC=0$ for 
$i\leq 3$ and equal to zero for $i>3$; and  for $\log(N_\rme)=27.5$ and
28.0 the carbon is fully ionized.

\subsection{The $W_n$ of {\rm OPAL}}
Bethe and Salpeter \cite{BS} give expressions for $\langle r_{n\ell}^3 \rangle$
for hydrogenic ions of charge $Z$. We put
\be 
\langle r_n^3\rangle=\left({1\over n^2}\right)
\sum_{\ell=0}^{n-1}(2\ell+1)\langle r_{n\ell}^3\rangle 
\ee
to obtain
\be 
\langle r_n^3\rangle=[n^2/(8Z^3)][21n^4+35n^2+4]\,. 
\ee
The mean volume for state $n$ can be defined as
\be 
V_n=(4\pi/3)\langle r_n^3\rangle\,. 
\ee

Let  $N_\rme$ be the electron density, $N_a$ be the atom density and $\rho$
be the mass density. From
Table 1 and Figure 1 of IR95, $N_\rme=5.0\times 10^{21}\rmcm^{-3}$ 
and, for $\log(\rho)=-2$
and the 6-element mix (see section \ref{mix6}), 
$N_a=4.7\times 10^{21}\rmcm^{-3}$. The
total particle density is then $N=N_\rme+N_a=9.7\times 10^{21}\rmcm^{-3}$.
The average volume occupied by a particle (electron or atom) is
$1/N$.

Table \ref{tabV} gives, for hydrogenic carbon and the case of Table 1
of IR95 ($\log(T)=6$, \mbox{$\log(\rho)=-2$}): values of
$NV_n$; $W_n$(OPAL) from IR95; and $W_n$(Q) from formulae given
in \cite{nay}. The values of $W_n$(Q) are fairly close to the values
of $W_n$(APEX), as given in IR95.

In both OP and OPAL, optical properties (oscillator strengths and
photoionization cross sections) are calculated for unperturbed
atomic states, and $W_n$ is the probability of state $n$ being
unperturbed. With $NV_n>1$, one would expect to find at least
one other plasma particle within the volume  $V_n$, giving a 
state which is markedly perturbed and which one would therefore expect 
to have a small value of $W_n$. By that criterion, some values of
$W_n$(OPAL) seem to be surprisingly large, particularly those
for $n=4$, 5 and 6. Values of $W_n$(Q) appear to be more reasonable.
\section{Inner-shell atomic physics}

The original OP work utilized the $R$-matrix method which
uses wavefunction expansions of the type
\be  
\Psi={\cal A}\sum_n\psi_n\theta_n\,,
\label{RM1}
\ee
where the $\psi_n$ are functions for atomic `target' states, the
$\theta_n$ are functions for an added electron, and $\cal A$ is an
anti-symmetrization operator. In that method, photoionization 
and autoionization are treated as a single quantum-mechanical process. 
Figure 2 gives, as an example, the cross section 
for transitions from the 1s2s $^1$S state of Fe$^{24+}$. Below the
threshold for ejection of the 1s electron, the cross section shows
autoionization features due to processes
\be 
1 \rms2 \rms + h\nu \rightarrow 2 \rms n \rmp \rightarrow 1 \rms+ \rme^-\,. 
\ee
However, use of the $R$-matrix method would not be practicable for the computation
of the atomic data required for the present work, for two reasons: (a) in many
cases, the number of channels $n$ required in equation (\ref{RM1}) would be 
prohibitive (our experience with the RmaX work on inner-shell X-ray processes
\cite{RmaX} shows that a few Li-  and Be-like ions are the most that
could be treated with a reasonable timescale); (b)
it would be difficult to allow for pressure-broadening of
the autoionization features. We therefore use a perturbative approach,
as implemented in the program {\sc autostructure} \cite{AS1, AS2}.

We note that Figure 2 shows, qualitatively, the effects of interference 
between autoionization features and the background continuum. However,
detailed quantitative studies  \cite{pind} show that,
on averaging over resonance profiles, this interference 
is a very small effect and it can safely be neglected for our purposes. 
This is  the independent processes approximation. In the time-reversed case, 
this corresponds to treating dielectronic and radiative recombination separately. 
The second approximation required by our perturbative approach is an isolated resonance
treatment of the autoionizing features.
The effect of interacting resonances  has been investigated \cite{pind} 
for the reverse process of dielectronic recombination, and it also can safely be 
neglected for our purposes.

\subsection{Photoexcitation}
The downward probability rate for a radiative transition from an upper
state $u$ to a lower state $\ell$ is given by
\be 
A^\rmr_{u\rightarrow \ell}=\frac{1}{g_u} \frac{4\omega^3}{3\hbar c^3} S_{\ell u} \,,
\ee
where $g_u$ is the statistical weight for the upper level, $c$ is the
speed of light, $\omega=2\pi\nu$ where $\nu$ is the photon frequency,
and $S_{\ell u}$ is the bound--bound line-strength, as defined in \cite{TAS}.
Eissner {\etal} \cite{SS} give expressions for $S_{\ell u}$ in multi-configuration
$LS$- and intermediate-coupling (their equations (115) and (117), respectively):
their code, {\sc superstructure}, gives numerical values for $S_{\ell u}$ in atomic units.
The code {\sc autostructure}, which incorporates {\sc superstructure},
 gives the following data: 
(a)  the line-centre frequency; 
(b) the rates $(g_u/g_\ell)A^\rmr_{u\rightarrow\ell}$;
(c)~$A^\rmr_u\equiv\sum_l A^\rmr_{u\rightarrow l}$, the total probability rate for radiative decay of
the upper level $u$; 
(d) $A^\rma_{u\rightarrow m}$ and $A^\rma_u\equiv\sum_m A^\rma_{u\rightarrow m}$,
the partial and total autoionization decay probability rates, respectively.
Multi-configuration $LS$- and intermediate-coupling expressions for 
$A^\rma_{u\rightarrow m}$ are given in \cite{AS1} (equations (2.2) and (2.4),
respectively).
The rates are all in s$^{-1}$.

The oscillator-strength for the $\ell\rightarrow u$ transition is 
\be 
f\!_{u\ell}=\frac{m_{\rm e}c^3}{2e^2\omega^2}\frac{g_u}{g_\ell}A^\rmr_{u\rightarrow \ell}
\ee
and the absorption (photoexcitation) cross section for an autoionizing feature
is 
\be 
\sigma^{\rm PE}_{\ell\rightarrow u}=\frac{2\pi^2e^2}{m_{\rm e}c}f\!_{u\ell}\varphi(\omega) \,,
\label{sigPE}
\ee
where $\varphi(\omega)$ is the {\em line-profile factor}. On neglecting
thermal Doppler broadening, the profile is
\be 
\varphi(\omega)=(\Gamma/2\pi)/[(\omega-\omega_0)^2+(\Gamma/2)^2] \,,
\label{phi}
\ee
where $\omega_0$ is the line-centre angular frequency and
\be 
\Gamma=(A^\rmr_\ell+A^\rmr_u)+A^\rma_u+\Gamma_{\ell u}^{\rm p} \,, 
\ee
where expressions for the pressure-broadening contribution, 
$\Gamma_{\ell u}^{\rm p}$, are given in \cite{adoc8}
(a different form is used for hydrogenic lines, see \cite{adoc13}). The profile
(\ref{phi}) is then convolved with that for Doppler broadening
to give a Voigt profile.

For more detailed (collisional--radiative) modelling purposes we require
to follow the break-up of an autoionizing state. For example, the photoexcitation
cross section in equation (\ref{sigPE}) is multiplied by branching
ratios for radiative decay, $A^\rmr_{u\rightarrow l}/(A^\rmr_u+A^\rma_u)$, and by Auger 
yields for autoionizing decay, $A^\rma_{u\rightarrow m}/(A^\rmr_u+A^\rma_u)$.

\subsection{Photoionization}
The direct photoionization cross-section from an initial state $\ell$
of an $(N+1)$-electron atom to a final state $u$ of an $N$-electron
ion plus ejected electron is given by \cite{ADOC1}
\be 
\sigma^{\rm PI}_{\ell\rightarrow u}=\frac{1}{g_\ell} \frac{4\pi^2}{3c}
\omega S_{u\ell}  \,,
\label{sigPI}
\ee
where $g_\ell$ is the statistical weight of the initial level
and $S_{u\ell}$ is the bound--free line-strength with the final-state continuum 
wavefunction normalized  per unit energy. The code {\sc autostructure} uses a different 
continuum normalization and gives numerical values for the bound--free line strength 
which are equal to $\pi/2$ times $S_{u\ell}$ in atomic units.
The same expressions for $S_{u\ell}$ that were used for bound--bound transitions are used
for bound--free transitions, except that the final `active'
bound-state wavefunction is replaced by a continuum distorted-wave
function which does not contain any resonance structure. This use of 
distorted-waves is a good approximation for atoms that are a few times ionized
and is the final approximation employed by our perturbative approach.

\subsection{{\sc autostructure}, some details}

\subsubsection{Angular algebra}
{\sc autostructure} \cite{AS1, AS2} incorporates {\sc superstructure} \cite{SS} and 
the angular algebra required to calculate the preceding atomic data is no more than
that which is generated by {\sc superstructure} to determine energy levels
and radiative rates. (The angular algebra required for autoionization rates
is that which follows from the $H$ operator which determines the structure.)
However, for the complex inner-shell processes considered here, which can give
rise to configurations with thousands of terms, we found it necessary to
re-write the angular algebra code. Specifically, because of historical memory
limitations, the algebraic Slater-state interactions were determined between each
$LS$ term or $J$ level \cite{SS}. However, the Slater-state interaction depends only trivially
on the configuration and can be generated much more efficiently between
symmetry groups. With complex configurations there is a high degree of algebraic 
term and level degeneracy, both within and between configurations. Re-coupling
by $LS$ or $LSJ$ symmetry groups reduces the overall time spent on the largest scale jobs
by a factor of 30--40 ($LS$) or 100--200 ($LSJ$).

\subsubsection{The Hamiltonian matrix}
{\sc superstructure} determines multi-configuration eigenenergies and eigenvectors 
by diagonalization of the Hamiltonian matrix for each $SL\pi$ or $J\pi$ symmetry group.
{\sc autostructure} further partitions the problem by $(N+1)$-electron bound and
$N$-electron (plus continuum) configurations --- the $N$- and $(N+1)$-electron
Hamiltonians are diagonalized separately. The bound--free Hamiltonian interaction 
between the two is treated as a perturbation --- this leads simply to the autoionization rate \cite{AS1}.

The mass--velocity and Darwin operators of the Breit--Pauli Hamiltonian \cite{SS}
are also added to the usual non-relativistic Hamiltonian for the
determination of our $LS$-coupling atomic structure. This results-in
transition energies between terms which are in good agreement with those
obtained from using intermediate coupling, on averaging-over fine-structure.
\subsection{Application to the 6-element mix}
The $K$-shell processes required are of the form:
\be 1\rms^q 2l^p n'l' +h\nu\rightarrow 1s^{q-1} 2l^p n'l' +\rme^- 
\label{K2}\ee
for photoionization and
\be 1\rms^q2l^pn'l'+h\nu\rightarrow 1\rms^{q-1}2 l^pn'l'n''l''
\rightarrow 1s^{q-1}2l^p n'''l''' +\rme^- \label{K1}\ee
for photoexcitation-autoionization, where
$2l^p$ stands for 2s$^s$2p$^t$ with $p=s+t$. 
Calculations are made for $q=1,2$ for $p=0$ and $q=2$ for $p>0$. 
The values of $p$ depend
on the element, for example, up to $p=7$ for iron but only $p=1$ for carbon. This
depends on the  importance of its contribution to the
opacity. We use $n', n'', n''' = 2$ to 6, for all allowed $l',l'',l'''$.
The contributions from higher-$n$ in (\ref{K1}) are obtained by matching onto
the results of (\ref{K2}).

The $L$-shell processes required are
\be 
2l^q 3l'^p n''l'' +h\nu  \rightarrow 2l^{q-1} 3l'^p n''l''+\rme^-
\label{L2} 
\ee
and
\be 
2l^q 3l'^p n''l'' +h\nu \rightarrow 2l^{q-1} 3l'^p n''l'' n'''l'''
\rightarrow 2l^{q-1} 3l'^p n^{iv}l^{iv} +\rme^- \,,
\label{L1} 
\ee
where  $3l'^p$ stands for for
3s$^s$3p$^t$3d$^u$ with $p=s+t+u$. Calculations are made for
$q=1$ to 8  for $p=0$ and $q=8$ for $p>0$; e.g., up to $p=2$ for iron.

The $M$-shell processes required are simpler:
\be 
3l^q n'l' + h\nu \rightarrow 3l^{q-1} n'l' +\rme^-
\label{M2} 
\ee
and
\be
3l^qn'l'+h\nu\rightarrow3l^{q-1}n'l'n''l''\rightarrow 3l^{q-1}n'''l'''+\rme^-\,,
\label{M1}
\ee
and are included only for iron, with $q=1$ to 3.

To attain as much accuracy as possible for consistency with the
existing $R$-matrix data, our perturbative calculations retain
configuration interaction within the $N$-electron (core) complex
and the ($N+1$)-electron complex (for $n'=n''$): for example, we retain 
interactions between states such as 2s$^2\,^1$S and 2p$^2\,^1$S.
For most cases, calculations are made both in $LS$-coupling and
in intermediate coupling (including Breit--Pauli terms).
Even when not significant for inner-shell
contributions to opacity, this resolution is required for modelling
non-LTE photoionized plasmas.

In table \ref{tabKLM}, we summarize the inner-shell calculations that we have
carried-out. We have examined the convergence with respect to the inclusion
of inner-shell data and estimate that the inclusion of further data would
change the Rosseland means by less than 1\%.

The new data resulted-in  an additional
2\,575\,458 level-resolved photoexcitation lines and 187\,351 
total photoionization cross sections, i.e. summed-over all final states,
contributing to the OP opacities. Much more data are
archived, viz. final-state
resolved photoionization cross sections and data for transitions and lines
than
were already included in the original OP work.

\subsection{Database issues}
Given that we have already generated, and will be generating, large amounts of 
atomic data, some thought has been given to how it might be effectively
archived for applications other than the one at hand. In particular, the
Atomic Data and Analysis Structure (ADAS) \cite{ADAS} has long had the
capability of handling radiation fields \cite{BS69} but has, until recently,
concentrated on utilizing escape factors \cite{FSH}. The collisional--radiative
modelling of finite-density non-LTE photoionized plasmas with ADAS requires
that we archive final-state resolved photoionization data, summed-over final 
channel angular momenta. 
We have written codes ({\sc adaspe} and {\sc adaspi}) to process the large amount 
of energy levels, radiative rates, autoionization rates and photoionization cross 
sections produced by {\sc autostructure} and have defined such a suitable archive 
data structure of final-state resolved photoexcitation-autoionization and direct 
photoionization data, specifically, {\it adf38} and {\it adf39} ADAS 
data formats \cite{ADAS}. This partial data is then further
reduced to total photoionization data for use by opacity calculations. 
For completeness, outer-shell photoionization data is archived in the {\it adf39} 
files as well as inner-shell, but it is not used in the work reported-on here.

\section{Free--free transitions}
Contributions from free--free transitions are, in most cases, calculated
in a hydrogenic approximation using the code of \cite{DGH} which employs 
fits to accurate results from \cite{KL}. Allowance for Debye screening
can be of importance at high density and is included if it gives 
increases in Rosseland-means by more than 0.1\%. The screening
contribution is calculated using the Born--Elwert theory (\cite {BS, Elwert, Roz}). 
It is checked that, without screening, the Born--Elwert
approximate gives agreement with the accurate results from \cite{DGH} and
\cite{KL} to better than 1\% for all cases considered in the present 
work. For $\log(R)\leq -1$ and $\log(T)\leq 7.7$ it is found that inclusion
of screening for free--free transitions never changes the Rosseland-mean
by more than 2 or 3 \%.

The correction for Debye screening was not included in the work of SYMP
but is included in the present work.
\section{Results for Rosseland-mean opacities}
\subsection{Use of the variable $\log(R)$}
It is convenient to use the variable
\be 
R=\rho/T_6^3 \,,
\ee
where $\rho$ is the mass density in g cm$^{-3}$ and $T_6=10^{-6}\times T$,
with $T$ in K. For a solar mix,
comparisons of $\log(\kappa_{\rm R})$ from OP and 
OPAL for $\log(R)=-1$ to $-6$ were given in Figure 15 of SYMP, which 
showed the OPAL
opacities to be larger than those from OP for larger values of
$\log(R)$ with $\log(T)>5.5$. 
\subsection{The 6-element mix\label{mix6}}
Iglesias and Rogers in IR95 considered the importance of 
inner-shell transitions for a mixture of 6 elements (H, He,
C, O, S and Fe) with abundances chosen to be such as to give
opacities similar to those for the complete solar mix. The
adopted number-fractions, $f_k$, are given in Table \ref{tabmix}. 

Paper IR95 gave results for one temperature--density point, $\log(T)=6$
and $\log(R)=-2$ giving $\log(\rho)=-2$. OPAL calculations were made both
with and without the inclusion of inner-shell processes. The inclusion of 
those processes was found to give an increase in $\kappa_{\rm R}$ by 30\%.
\subsection{{\rm OP} results both with and without inner-shell transitions}
Figure 3 gives OP values of $\log(\kappa_{\rm R})$ for the 6-element mix,
both with the inclusion of the inner-shell data discussed in Section 3
and without. The results without inner-shell data are essentially the 
same as those given in SYMP, but include some fairly minor improvements and 
use of the Q-EOS. Figure 4 shows 
$\delta\log(\kappa_{\rm R})$, the increase in $\log(\kappa_{\rm R})$ due
to inclusion of the inner-shell data. For $\log(T)=6,\,\log(R)=-2$ we
obtain an increase in $\kappa_{\rm R}$ by 31\%, in close agreement with 
the increase obtained in IR95.

Changes in $\kappa_{\rm R}$ due to changes in the EOS are much smaller
than those due to the inclusion of inner-shell data: thus for $\log(T)=6$, 
$\log(R)=-2$, use  of the Q-EOS in place of the EOS used in SYMP
reduces $\kappa_{\rm  R}$ by only 1.6\% (in both cases, without inner-shell
data).
\subsection{Results from {\rm OP} and {\rm OPAL}}
Figure 5 shows values of $\log(\kappa_{\rm R})$ for the 6-element mix from 
OPAL (data obtained from the OPAL website \cite{web}) and from the present  
OP work with the inclusion of inner-shell data.  It is seen that, for 
the larger values of $\log(R)$ where the inner-shell data is important,
OP is now in close agreement with OPAL.

There are some remaining differences between OP and OPAL at smaller
values of $\log(T)$ and $\log(R)$ where inner-shell data are not important.
A feature at $\log(T)\simeq 5.2$, often referred to as the `$Z$-bump', was 
first identified by Iglesias, Rogers and Wilson \cite{irw87} as due to 
inclusion of very
large numbers of $M$-shell transitions for various ionization stages of 
iron. For $\log(R)\leq -3$ there are seen to be some differences
between OPAL and OP in the vicinity of that feature. Figure 6 shows, on a
much more expanded scale, percentage difference for $\kappa_{\rm R}$, 
(OP$-$OPAL), for  $\log(R)=-3$ and $-6$. For $\log(R)=-3$ the $Z$-bump feature
from OP is seen to be broader than that from OPAL, and to have a peak value
lower by about 10\%. For $\log(R)=-6$ the OP peak value is lower by 20\%.
We plan to consider these differences further in a later paper.
\section{Discussion}
IR95 was mainly concerned with illustrative results for the case
of the 6-element mix and $\log(T)=6$, $\log(\rho)=-2$.
Plots of monochromatic opacities were given for C, S, Fe and for the mixture.
Similar plots were obtained in the course of the present work.
\subsection{Density dependence}
It is seen from Figure 4 that the inner-shell contributions are
most important at high densities. That can be understood by considering
the case of high temperatures (similar considerations
apply for other regions). At high temperatures, say $\log(T)\geq 7.0$,
the most important transitions are for iron $K$-shell. At the very lowest 
densities considered, the iron is almost fully ionized and there are no
$K$-shell contributions. Then, as the density increases,
one has $K$-shell transitions of the type
\be 
1\rms+h\nu\rightarrow \kappa\rmp\,,
\ee
and
\be 
1\rms^2+h\nu\rightarrow 1\rms\kappa\rmp\,,
\ee
where $\kappa=n$ for transitions to a bound state and $\kappa=k$
for photoionization. Such transitions are included in the original
OP calculations. With a further increase in density, one has states
of the type 1s$^2C$ (where `$C$' stands for states of outer electrons) 
and $K$-shell transitions of the type
\be 
1\rms^2C+h\nu\rightarrow 1\rms C \kappa\rmp\,.
\ee
Such transitions were 
not included in the original OP work but are included in the
present work (see section 3).
\subsection{Line profiles}
Pressure broadening of all spectrum lines should be included, for
both transitions to true bound states and for those to autoionizing
states.

The importance of pressure-broadening for the iron $K$-shell is
illustrated in Figure 7, which gives plots of $\sigma(u)$ for
iron for $\log(T)=7$ and $\log(N_\rme)=22.5$ and 25: for the 6-element
mix, those values of $N_\rme$ correspond to $\log(R)=-4.21$ and $-1.71$.
It is seen that, for the low-density case, there are very many
resolved spectrum lines but that at higher densities pressure broadening
leads to the lines being almost completely blended. We use the 
pressure-broadening theory from \cite{adoc8} and \cite{adoc13}
and line-blending theory from \cite{dam}.

We note that, beyond the $K$-edge, the cross sections are nearly
the same for both densities, since both depend mainly on
$K$-shell ionization. At the lower energies, say $u\leq 7$, the
background opacity is mainly due to free--free transitions, 
giving cross-sections per atom proportional to the electron density.
\subsection{Autoionization}
The {\sc autostructure} calculations give line widths due to
radiation damping and to autoionization. We find, in practice,
that inclusion of those contributions to the line profiles
is never very important since widths due to pressure broadening 
are generally much larger:
omission of autoionization widths
would never give errors in Rosseland-means larger than one
on two percent. 
\subsection{Fine-structure}
In both the OPAL and OP work, it was found that inclusion of 
fine-structure for outer-shell transitions could be of importance
for the calculation of Rosseland-means (see \cite{irw92} and SYMP).
Most of our {\sc autostructure} calculations were made both in
$LS$-coupling (no fine-structure) and in intermediate coupling
(with fine-structure). Test runs showed that inclusion of fine-structure
for the inner-shell transitions never increased Rosseland-means by
as much as 1\%, which was, again, a consequence of the importance
of pressure-broadening. All final results as reported in the present
paper were made with omission of inner-shell fine-structure.
\section{The solar centre region}
Convection occurs throughout much of the solar interior and in those
regions a precise knowledge of the Rosseland-mean opacity is not
of great importance. However, convection does not occur
in the deepest layers of the solar interior. Knowledge of the Rosseland-mean
for the centre region is of importance for the construction
of solar models, which can be tested against data from helioseismology.

At the solar centre, models give $\log(T)=7.196$, $\log(\rho)=2.179$ 
and $\log(R)=-1.409$ (see \cite{GDKP,BU}). Opacities
calculated using a number of different codes \cite{rose} show quite
a large scatter. The most accurate values currently available are
undoubtedly those from OPAL \cite{IR91}. For the 6-element mix, opacities
for the centre region from OP are a little larger than those from OPAL:
by 2.3\% at $\log(T)=7.2$ and $\log(R)=-1.5$.

The present work permits some further discussion of the solar-centre problem.
In the OP work we use a mesh of values of $\log(T)$ and $\log(N_\rme)$.
We select a mesh-point $\log(T)=7.2$, $\log(N_\rme)=26$ giving, for
the 6-element mix, $\log(\rho)=2.294$ and $\log(R)=-1.306$. Figure 8
shows $\log(\sigma)$ for the mixture at that point. There are three
main contributions to the centre opacity.
\subsection{Free--free} 
The free--free contribution is mainly due to electron collisions with 
H$^+$ and He$^{+2}$.
The free--free cross section behaves like $u^{-2}$ and the process
therefore dominates at low frequencies. We include Debye screening
(see Section 4) which modifies the solar-centre opacities by
about 1 or 2 per cent. OPAL include some further refinements (see \cite{IR95}) but
they are not likely to be of much importance.
\subsection{Electron scattering} 
At lower densities, the electron-scattering
cross section is equal to the Thomson cross section, and is independent
of frequency. For higher densities, the cross section is modified by
plasma collective effects.
Both OP and OPAL use the theory of Boercker \cite{brk}, and also allow
for relativistic corrections.
\subsection{Atomic transitions}
In Figure 8 there is a 
just-discernible feature at $u\simeq 2$ due to sulphur $K$-shell transitions, and
a much more conspicuous one at $u\simeq 5$ due to iron $K$-shell transitions.
For hydrogenic iron, the Ly$_\alpha$ line is at $u=5.10$ and the Lyman
continuum starts at $u=6.80$. Table 3 gives ionization fractions
and ground-state occupation probabilities for iron: all stages up to
$i=9$ are seen to contribute to the $K$-feature. It is seen from Figure 7 that the 
high $K$-shell lines are completely blended. It may be noted that the
mean opacity will not be sensitive to the exact distribution amongst
ionization stages (the $\phi_i$
of Table \ref{tabfrac}) since the cross
sections for promotion of 1s$^2$ electrons will be much the same
for the different stages.

\subsection{Use of a different frequency variable}
Equation (4) may be replaced by
\be \frac{1}{\sigma_{\rm R}}=\int_{v=0}^{v_{\rm max}}
\frac{1}{\sigma(u)}\rmd v\, \label{eqv} 
\ee
where
\be 
v(u)=\int_{u=0}^v F(u) \rmd u 
\ee
and $v_{\rm max}=v(u\rightarrow\infty)$ --- numerical integrations give 
$v_{\rm max}=1.0553$. Figure 9 shows 
$1/\sigma$ plotted against $v$ for the case of Figure 7. The advantage
of using Figure 9 is that it shows the sensitivity of $1/\sigma_{\rm R}$ to 
the various features in $\sigma(u)$. There are seen to be rather small
contributions from regions of small $u$ where $\sigma(u)$ is large but much
more important contributions from large $u$ where $\sigma(u)$ is small.
\section{Summary}
Rosseland-mean opacities $\kappa_{\rm R}$ from the Opacity Project, OP, were
originally found to be smaller than those from the OPAL project at high temperatures
and high densities. Iglesias and Rogers, in IR95 \cite{IR95}, discussed the case
of $\log(T)=6$, $\log(\rho)=-2$ where $\kappa_{\rm R}$(OPAL) was larger than
$\kappa_{\rm R}$(OP) by about 30\%. They made two criticisms of the OP
work: (a) OP calculated occupation probabilities $W$ using a Holtsmark 
MFD where,  at high densities, it is a poor approximation;
(b) OP omitted some important inner-shell atomic data.
\subsection{Occupation probabilities} 
In the present work we calculated values of $W$ using expressions from
Nayfonov \etal \cite{nay}, who used an MFD theory that is valid to high 
densities. It is shown that the expressions for $W$ given in \cite{hm} and
\cite{nay} are not valid in the limit of very high densities: a simple
expedient removes that difficulty. It is found that final results for
Rosseland-mean opacities are not very sensitive to the adopted 
occupation probabilities.

It is noted that the values of $W$ obtained from the OPAL work
seem to be rather surprisingly large for more highly-excited states.
\subsection{Inner-shell data}
New opacity calculations have been made for the 6-element mix introduced in
IR95. Extensive inner-shell atomic data were computed using the code
{\sc autostructure} for the 6-element mix introduced in IR95. It is shown 
that inclusion of those data removes all major differencies between the
OP and OPAL work.
\subsection{Future work}
In the work described in SYMP, opacities were calculated for 17
cosmically-abundant chemical elements. Work is now in progress to
obtain inner-shell data for those elements not included in the
present work.
\newpage
\section*{References}
\newpage
\begin{table}
\caption{Occupation probabilities ($W_n$) and 
\newline
average volumes ($V_n$) for C$^{5+}$ at 
\newline
$T=10^6$K and $N=9.7\times10^{21}\rmcm^{-3}\,.$\label{tabV}}
\footnotesize\rm
\begin{indented}
\item[]\begin{tabular}{llll}
\br
$n$  & $NV_n$ & $W_n$(OPAL) & $W_n$(Q)\\
\mr
$1$&$2.08(-4)^a$&$1.000  $&$1.000  $\\
$2$&$6.67(-3)$&$0.996  $&$0.997  $\\
$3$&$6.31(-2)$&$0.995  $&$0.967  $\\
$4$&$0.330  $&$0.995  $&$0.705  $\\
$5$&$1.216  $&$0.914  $&$0.154  $\\
$6$&$3.562  $&$0.527  $&$1.58(-2) $\\
$7$&$8.875  $&$0.162  $&$1.87(-3) $\\
$8$&$1.96(+1) $&$2.37(-2)$&$2.82(-4) $\\
$9$&$3.96(+1) $&$2.23(-3)$&$5.22(-5) $\\
\br
\end{tabular}
\item[]$^{\rma}2.08(-4)=2.08\times 10^{-4}\,.$
\end{indented}
\end{table}
\begin{table}
\caption{Inner-shell transitions considered: 
\newline
$Z$ is the nuclear charge of the element and $N$ the  
\newline
number of electrons left on the final ion.
\label{tabKLM}}
\footnotesize\rm
\begin{indented}
\item[]\begin{tabular}{rrrrrrrrrr}
\br
 $Z$ & $N$ & \multicolumn{2}{c}{Shells} && $Z$ & $N$ & \multicolumn{2}{c}{Shells} \\
\mr
  2 & 1 & $ K$ &     &     	     &	  &     &     &     & \\
    &	&      &     &     	     &	  &     &     &     & \\
  6 & 1 & $K$  &     &               & 26 & 1   & $K$ &     & \\
    & 2 & $K$  &     &               &    & 2   & $K$ &     & \\
    & 3 & $K$  & $L$ &               &    & 3   & $K$ & $L$ & \\
    &	&  &   &                     &    & 4   & $K$ & $L$ & \\
  8 & 1 & $K$  &     &               &    & 5   & $K$ & $L$ & \\
    & 2 & $K$  &     &               &    & 6   & $K$ & $L$ & \\
    & 3 & $K$  & $L$ &               &    & 7   & $K$ & $L$ & \\
    &	&  &   &                     &    & 8   & $K$ & $L$ & \\
 16 & 1 & $K$  &     &               &	  & 9   & $K$ & $L$ &     \\
    & 2 & $K$  &     &               &    & 10  &     & $L$ &     \\
    & 3 & $K$  & $L$ &               &    & 11  &     & $L$ & $M$ \\
    & 4 & $K$  & $L$ &               &    & 12  &     & $L$ & $M$ \\
    & 5 &      & $L$ &               &    & 13  &     &     & $M$ \\
    & 6 &      & $L$ &     &         &    &     &     &   \\
    & 7 &      & $L$ &     &	     &	  &     &     &    \\
    & 8 &      & $L$ &     &	     &	  &     &     &    \\
    & 9 &      & $L$ &     &	     &	  &     &     &    \\
    &10 &      & $L$ &     &	     &	  &     &     &    \\

\br
\endTable
\clearpage

\begin{table}
\caption{Number fractions ($f_k$)  
\newline
for the 6-element ($k$) mix of 
\newline
Iglesias and Rogers \cite{IR95}.\label{tabmix}}
\footnotesize\rm
\begin{indented}
\item[]\begin{tabular}{lll}
\br
$k$ &  $f_k$\\
\mr
H    &$9.071(-1)$\\
He   &$9.137(-2)$\\
C    &$4.859(-4)$\\
O    &$9.503(-4)$\\
S    &$9.526(-5)$\\
Fe   &$3.632(-5)$\\
\br
\endTable
\begin{table}
\caption{Ionization fractions ($\phi_i$) and \\
ground-state occupation  probabilities ($W_0$)\\
for iron for $\log(T)=7.2$ and $\log(N_\rme)=26$.\label{tabfrac}}
\footnotesize\rm
\begin{indented}
\item[]\begin{tabular}{lllll}
\br
$i$    & $\phi_i$  &  $W_0$\\
\mr
$0   $  &$0.000 $    & $1.000$\\
$1   $  &$0.001 $    & $1.000$\\
$2   $  &$0.039 $    & $1.000$\\
$3   $  &$0.155 $    & $0.959$\\
$4   $  &$0.289 $    & $0.954$\\
$5   $  &$0.282 $    & $0.941$\\
$6   $  &$0.162 $    & $0.930$\\
$7   $  &$0.058 $    & $0.918$\\
$8   $  &$0.012 $    & $0.894$\\
$9   $  &$0.002 $    & $0.872$\\
$10  $  &$0.000 $    & $0.847$\\
\br
\endTable
%
%
\Figures
\begin{flushleft}
\epsfig{file=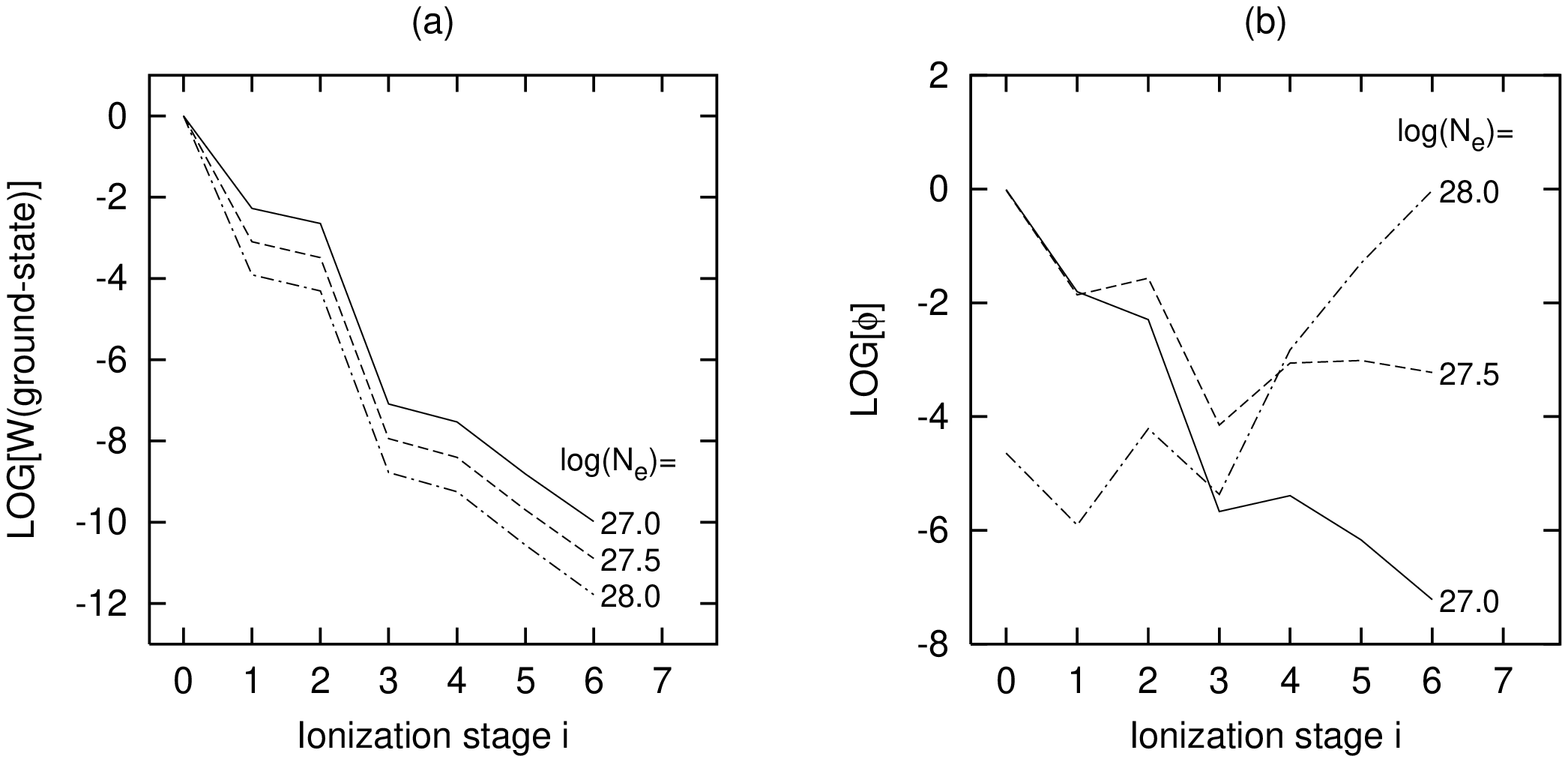,angle=270,width=14cm}
\end{flushleft}
\Figure{Carbon at $\log(T)=7.5$ and $\log(N_\rme)=27.0$, 27.5 and
28.0. (a) Ground-state occupation probabilities, $W$, against ionization
stage $i$. (b) Ionization fractions, $\phi_i$, calculated without a cut-off,
$W_\rmC$, in $W$.}
\newpage
\begin{flushleft}
\epsfig{file=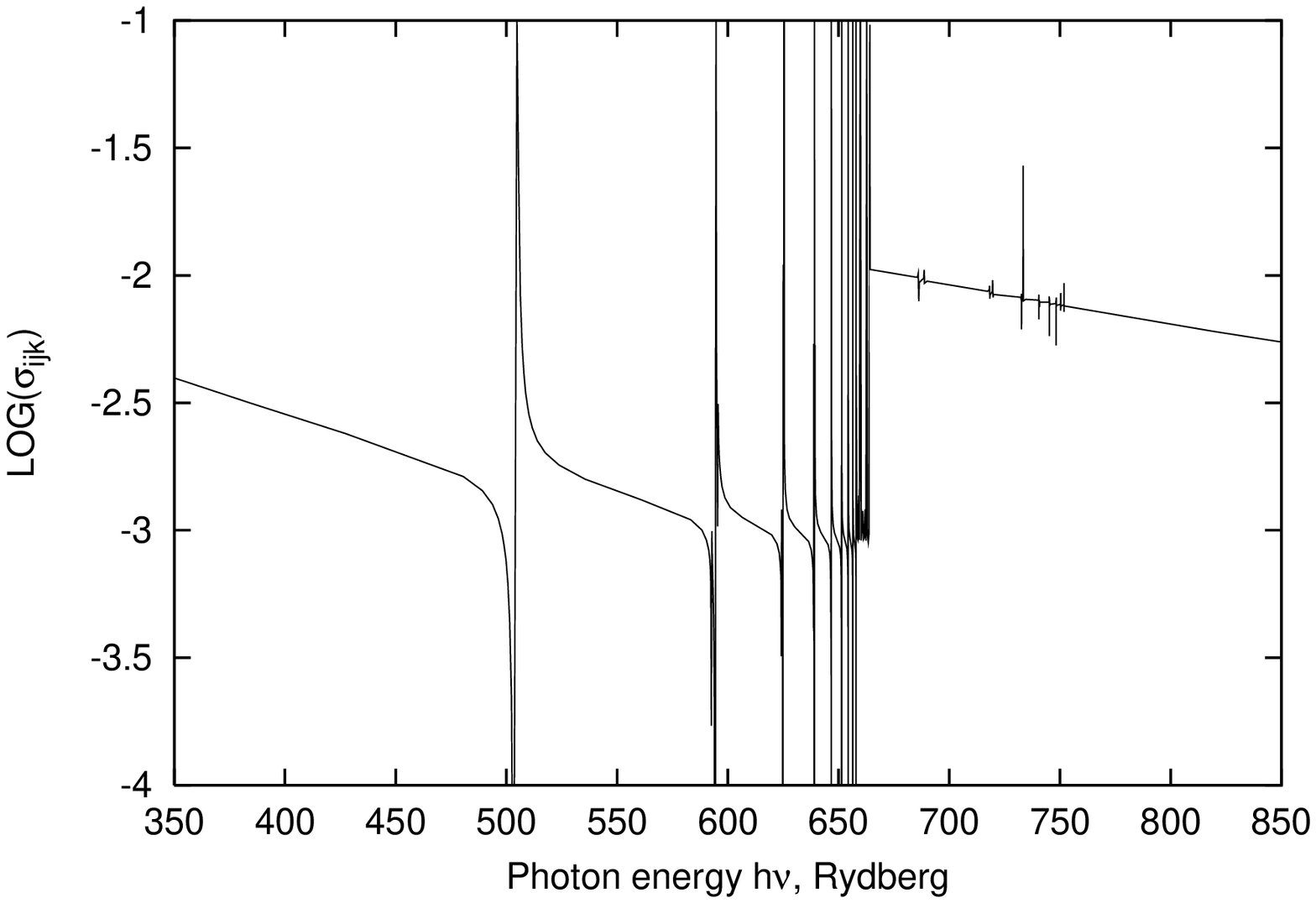,angle=270,width=12cm}
\end{flushleft}
\Figure{The atomic cross-section, $\sigma_{ijk}$, for photoionization 
from Fe$^{24+}$ 1s2s~$^1$S, in the
vicinity of the $K$-edge. Cross section in atomic units.}
\clearpage
\begin{flushleft}
\epsfig{file=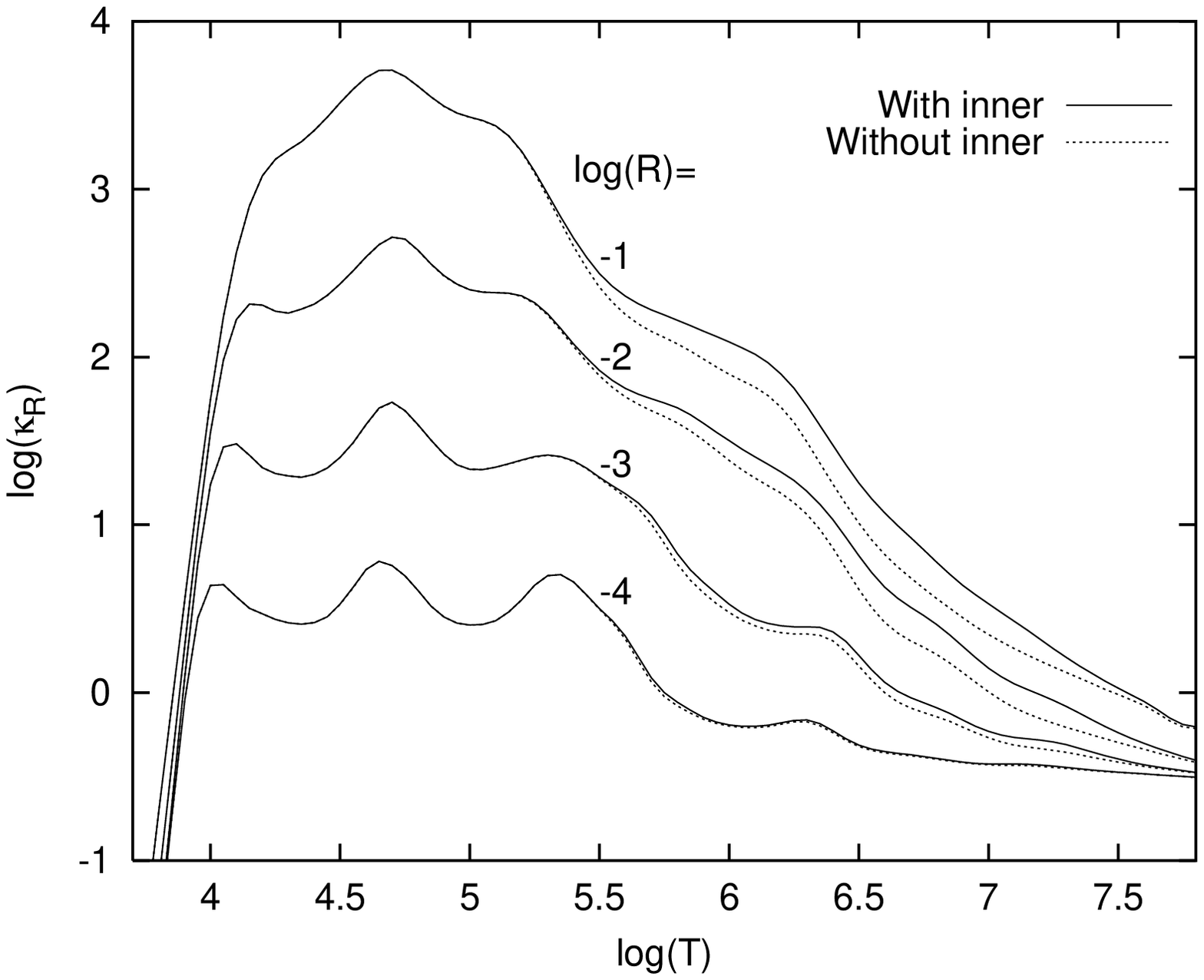,angle=270,width=14cm}
\end{flushleft}
\Figure{OP Rosseland-mean opacities for the 6-element mix for $\log(R)=-1$, 
$-2$, $-3$ and $-4$, with inclusion of inner-shell data (full lines) and without
those data (dotted lines). The opacities are in cgs units: cm$^2$ g$^{-1}$.}
\newpage
\begin{flushleft}
\epsfig{file=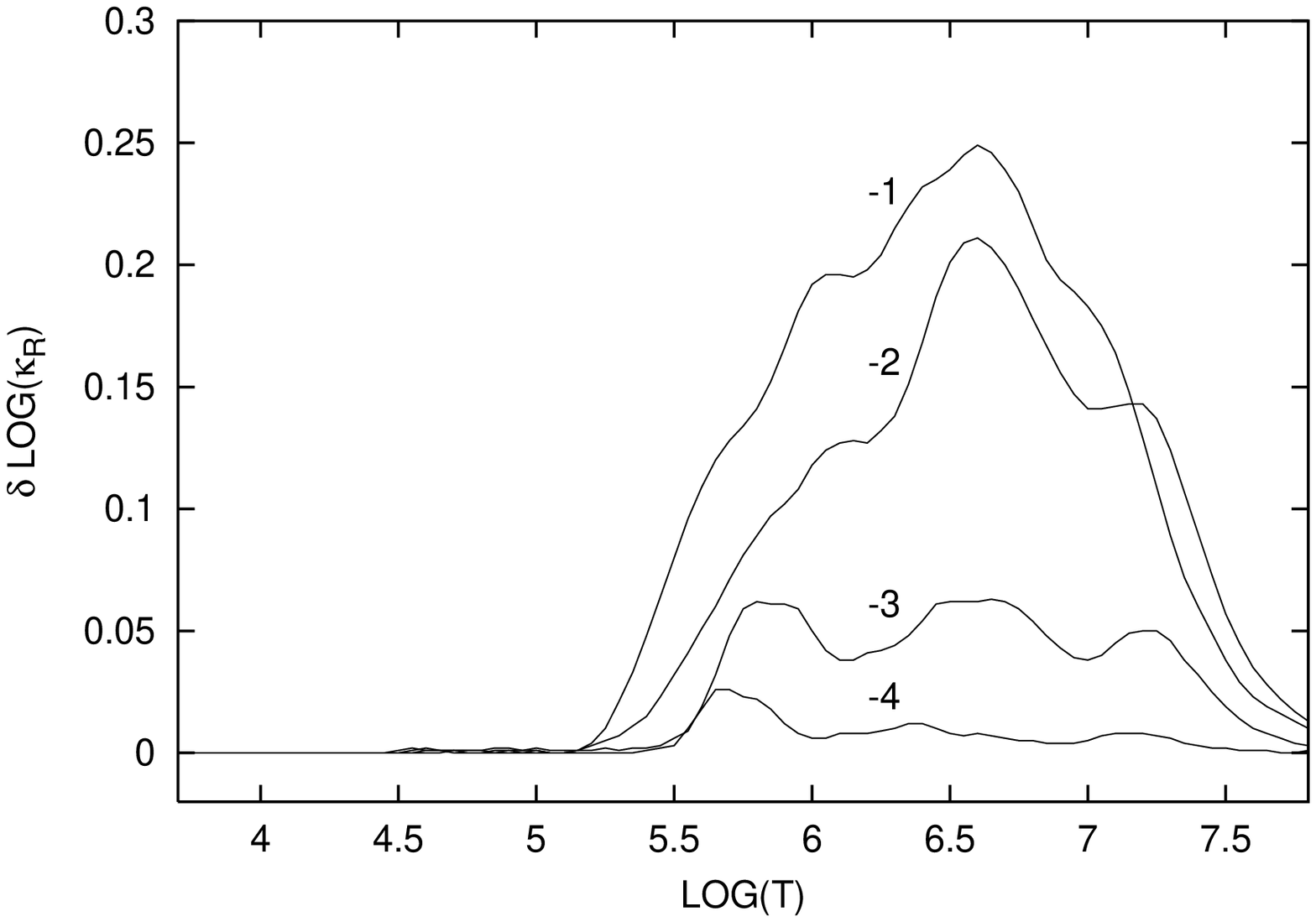,angle=270,width=12cm}
\end{flushleft}
\Figure{$\delta\log(\kappa_{\rm R})$, the change in Rosseland-mean
opacity which results from the inclusion of inner-shell data.}
\newpage
\begin{flushleft}
\epsfig{file=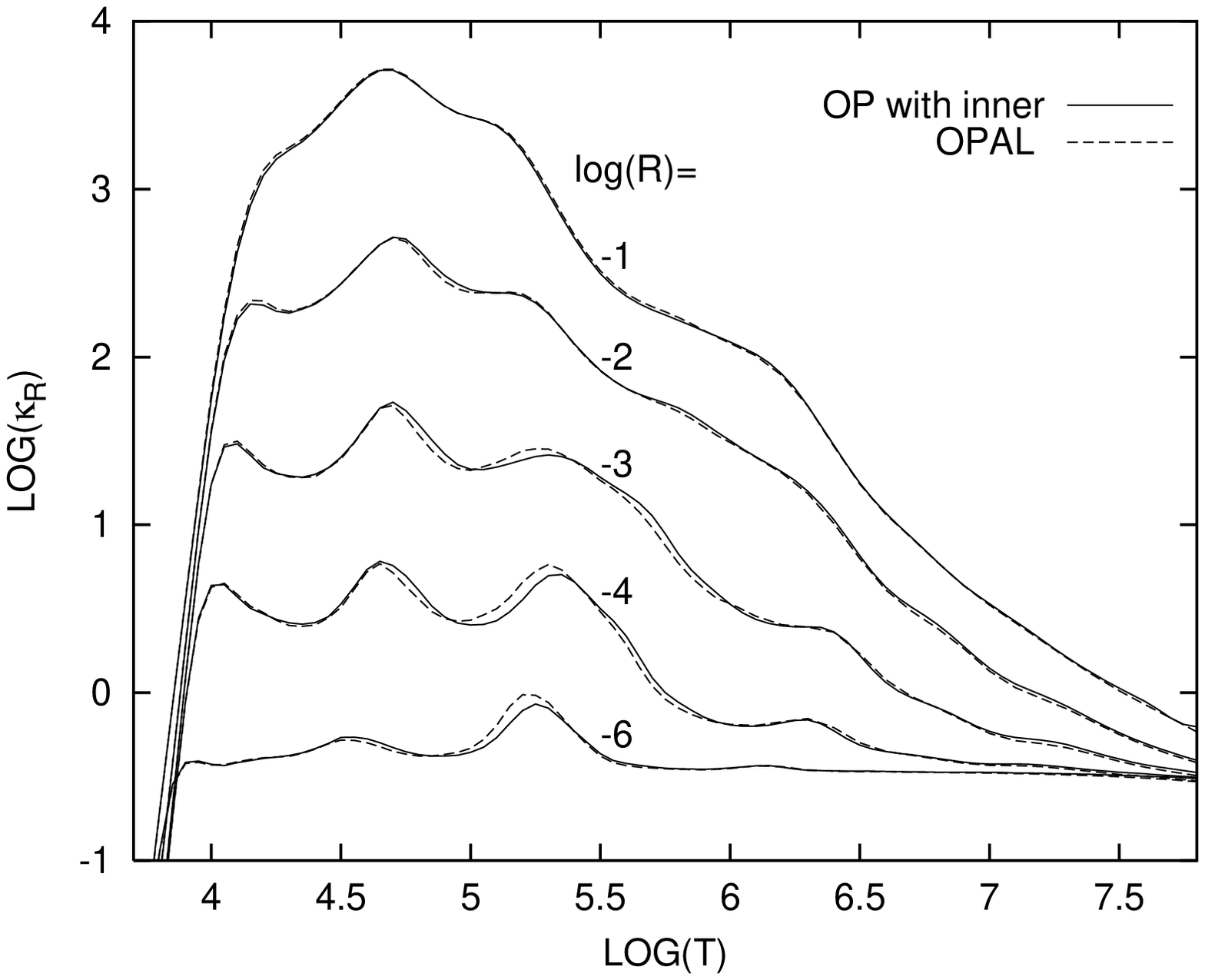,angle=270,width=14cm}
\end{flushleft}
\Figure{Rosseland-mean opacities for the 6-element mix, with the inclusion of 
inner-shell data: full lines, OP, present work; dashed lines, OPAL, from 
\cite{web}. Opacities in cgs units.}
\newpage
\begin{flushleft}
\epsfig{file=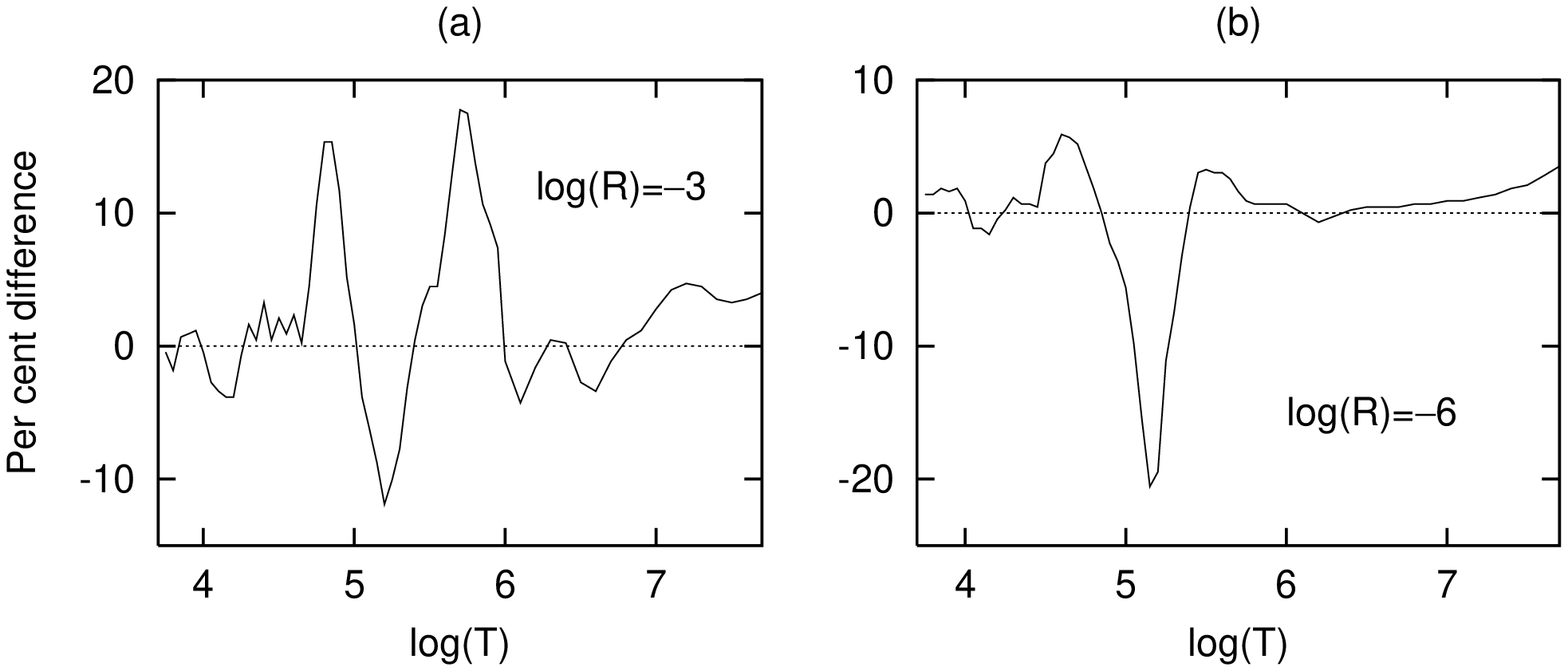,angle=270,width=15cm}
\end{flushleft}
\Figure{Percentage differences between OP and OPAL Rosseland means,
OP$-$OPAL. (a), $\log(R)=-3$. (b), $\log(R)=-6$.}
\newpage
\begin{flushleft}
\epsfig{file=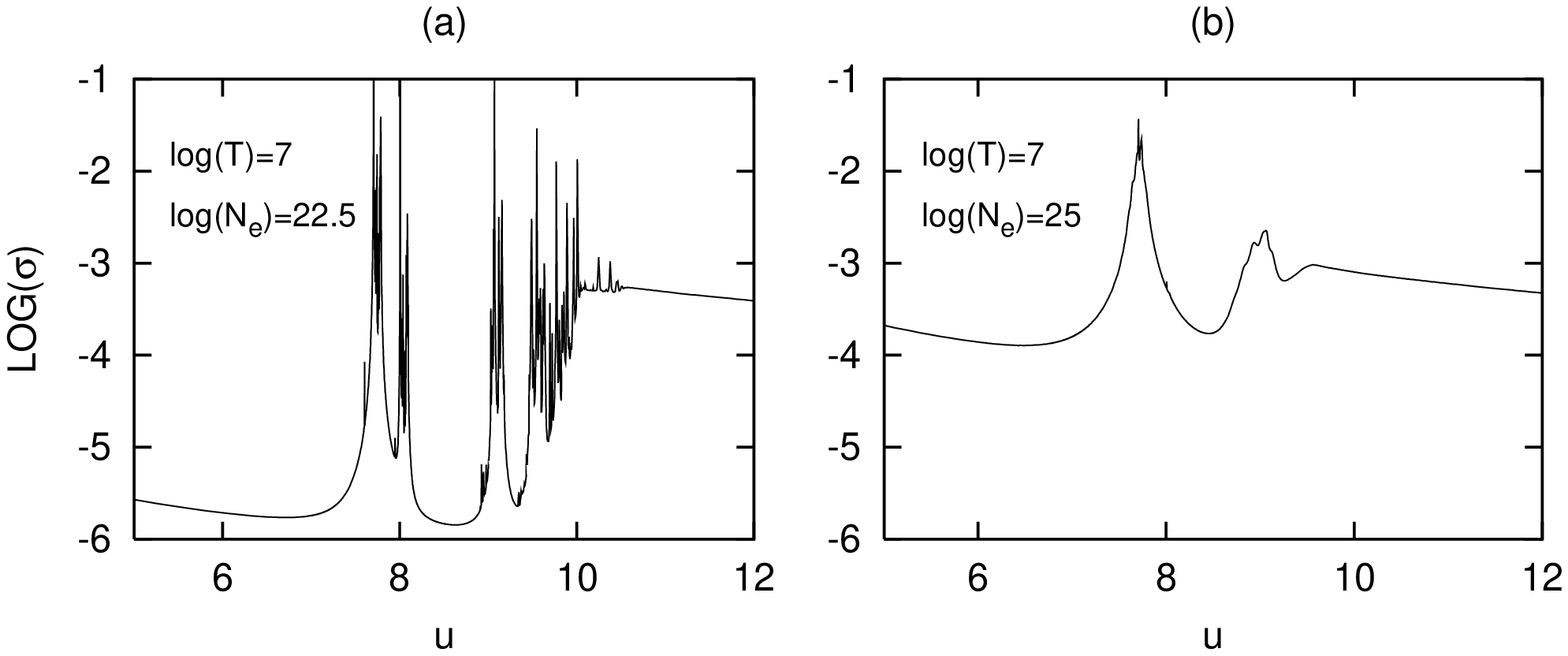,angle=270,width=15cm}
\end{flushleft}
\Figure{The opacity cross-section, $\sigma$, for iron in the vicinity
of the $K$-edge, at $\log(T)=7$: cross section in atomic units. Pressure
broadening included.\\
(a) $\log(N_\rme)=22.5$,  and (b) $\log(N_\rme)=25$.}
\newpage
\begin{flushleft}
\epsfig{file=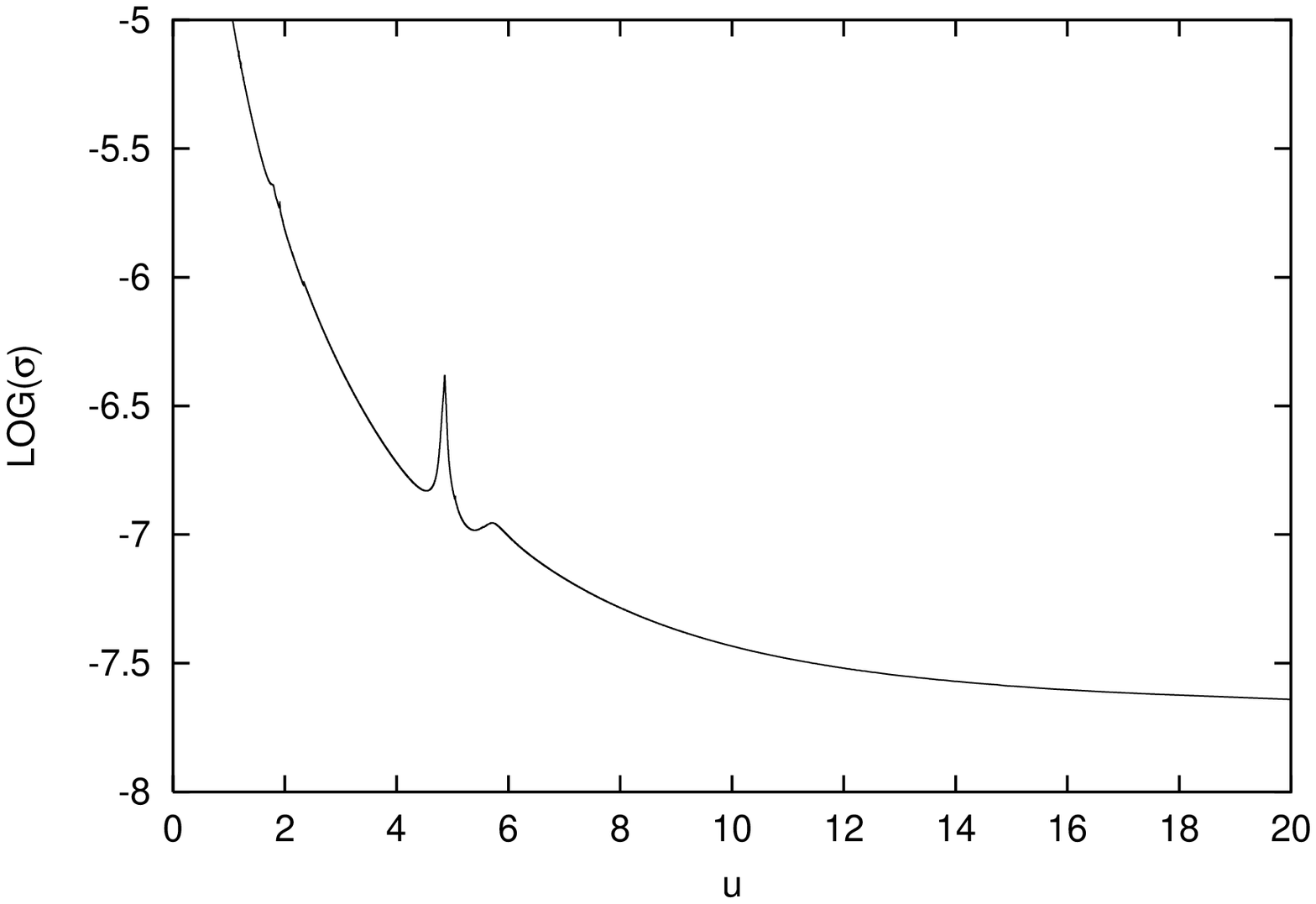,angle=270,width=12cm}
\end{flushleft}
\Figure{The opacity cross-section, $\sigma$, (in atomic units) for
the 6-element mix at $\log(T)=7.2$, $\log(N_\rme)=26$, $\log(R)=-1.306$.}
\newpage
\begin{flushleft}
\epsfig{file=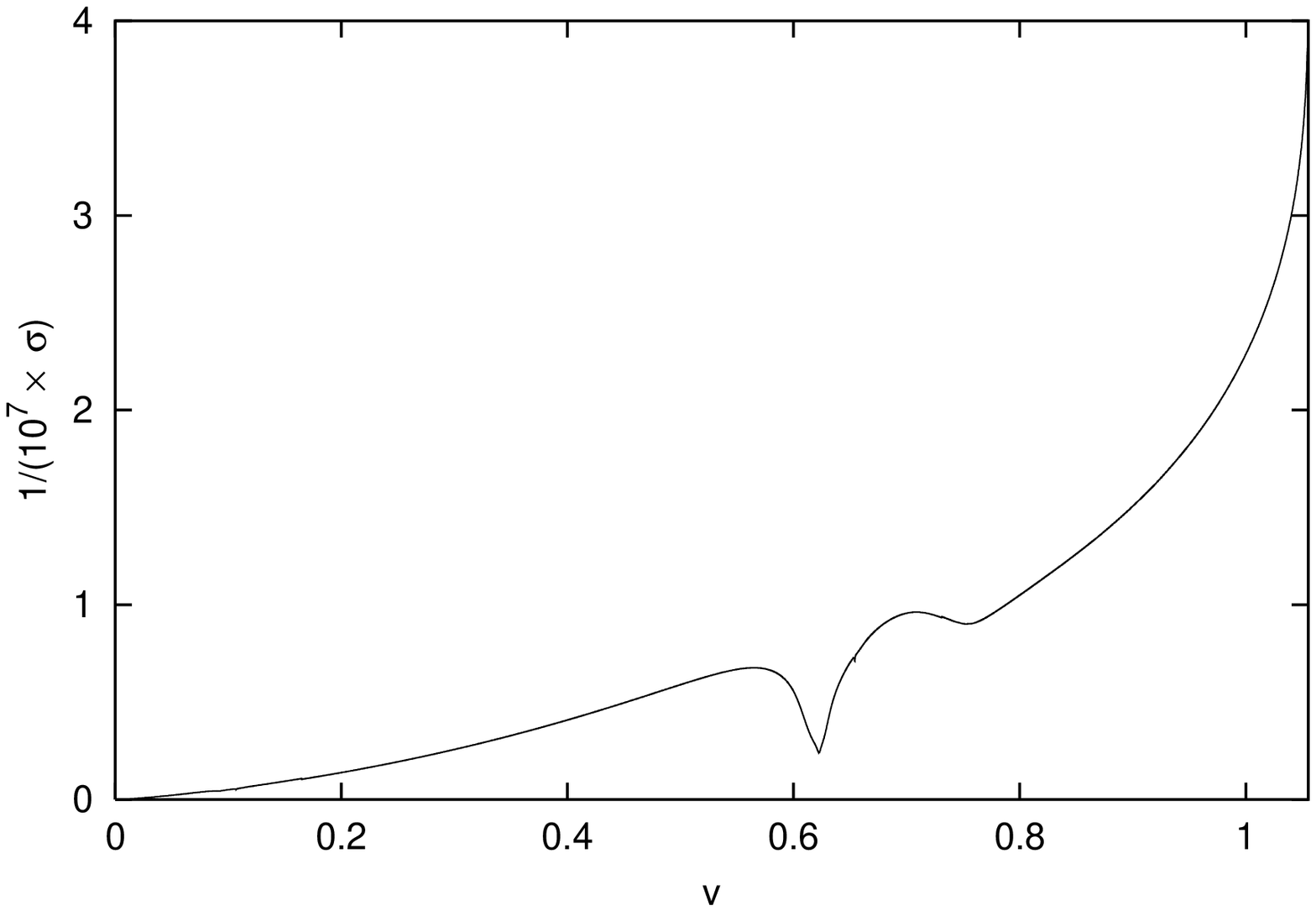,angle=270,width=12cm}
\end{flushleft}
\Figure{$1/\sigma$ against $v$, defined by equation (\ref{eqv}), for the case
of Figure 7.}

\end{document}